\documentstyle[12pt]{article} 

\newlength{\extraspace}
\newlength{\extraspaces}
\newcommand{\bQ}{{\bar{Q}}}

\newcommand{\PF}{{P_{\! F}}}

\newcommand{\bS}{{\bar{S}}}
\newcommand{\bQP}{{\bQ_{\! A}}}
\newcommand{\bQN}{{\bQ_{\! B}}}

\newcommand{\bQF}{{\bQ_{\! F}}}

\newcommand{\bBox}{{\Box^{\ast}}}
\newcommand{\iibBox}{{{\Box\;\!\!\!\Box}^{\ast}}}
\newcommand{\iiBox}{{{\Box\;\!\!\!\Box}}}
\newcommand{\bq}{{\bar{q}}}

\newcommand{\tN}{{\tilde{N}}}
\newcommand{\tQ}{{\tilde{Q}}}
\newcommand{\tW}{{\tilde{W}}}
\newcommand{\Qf}{{Q_{\! f}}}
\newcommand{\bQA}{{\bQ_{\! A}}}
\newcommand{\bA}{{\bar{A}}}

\newcommand{\bqP}{{\bar{q}_A}}
\newcommand{\cN}{{\cal N}}
\newcommand{\p}{{\partial}}

\newdimen\tableauside\tableauside=1.3ex
\newdimen\tableaurule\tableaurule=0.4pt
\newdimen\tableaustep
\def\phantomhrule#1{\hbox{\vbox to0pt{\hrule height\tableaurule
width#1\vss}}}
\def\phantomvrule#1{\vbox{\hbox to0pt{\vrule width\tableaurule
height#1\hss}}}
\def\sqr{\vbox{%
  \phantomhrule\tableaustep
\hbox{\phantomvrule\tableaustep\kern
      \tableaustep\phantomvrule\tableaustep}
  \hbox{\vbox{\phantomhrule\tableauside}\kern-\tableaurule}}}
\def\squares#1{\hbox{\count0=#1\noindent\loop\sqr
  \advance\count0 by-1 \ifnum\count0>0\repeat}}
\def\tableau#1{\vcenter{\offinterlineskip
  \tableaustep=\tableauside\advance\tableaustep by-\tableaurule
  \kern\normallineskip\hbox
    {\kern\normallineskip\vbox
      {\gettableau#1 0 }%
     \kern\normallineskip\kern\tableaurule}%
  \kern\normallineskip\kern\tableaurule}}
\def\gettableau#1 {\ifnum#1=0\let\next=\null\else
  \squares{#1}\let\next=\gettableau\fi\next}

\catcode`\@=11

\def\numberbysection{\@addtoreset{equation}{section}
\def\theequation{\arabic{section}.\arabic{equation}}}
\begin{document}

\thispagestyle{empty}

\begin{center}
\begin{flushright}
NHCU-HEP-97-15 \\
October, 1997 \\
\end{flushright}
\vspace{2.5cm}
\begin{center}

{\Large
{\bf On the $Z_2$ Monopole of $Spin(10)$ Gauge Theories}} \\[18mm]
{\sc Wang-Chang Su} \\[4mm]
{\it Department of Physics \\[2mm]
National Tsing-Hua University \\[2mm]
Hsinchu 300, Taiwan } \\[2mm]
{\tt suw@phys.nthu.edu.tw} \\[4mm]

\end{center}
\vspace{10mm}
%
{\parbox{13cm}
{\hspace{5mm}
An ``expanded" description is introduced to examine the 
spinor-monopole identification proposed by Strassler for 
four-dimensional $\cN=1$ supersymmetric $Spin(10)$ gauge 
theories with matter in $F$ vector and $N$ spinor representations. 
It is shown that a $Z_2$ monopole in the ``expanded" theory is 
associated with massive spinors of the $Spin(10)$ theory. 
For $N=2$, two spinor case, we confirm this identification 
by matching the transformation properties of the two theories 
under $SU(2)$ flavor symmetry. 
However, for $N \ge 3$, the transformation properties are not 
matched between the spinors and the monopole. 
This disagreement might be due to the fact that the $SU(N)$ 
flavor symmetry of the $Spin(10)$ theory is partially realized 
as an $SU(2)$ symmetry in the ``expanded" theory. 

\vspace{2mm}


}}

\end{center}

\vfill

\newpage


\section{Introduction}

The recent years have witnessed significant progress in the study 
of strongly coupled dynamics of four-dimensional $\cN=1$ 
supersymmetric gauge theories \cite{susy}. 
See \cite{recent} for recent reviews and references therein 
for earlier work. 
Most of this progress is attributed to the pioneering work 
of N. Seiberg \cite{seiberg} on the outstanding idea of duality: 
a SUSY gauge theory that is in a non-Abelian Coulomb phase 
may have an equivalent description in terms of a different 
gauge groups and matter content. 
A strongly coupled SUSY gauge theory thus has a weakly coupled dual 
that allows the theory to be solved. 
Thanks to Seiberg's work, many examples of duality involving 
complicated gauge group and matter content have been 
uncovered \cite{fundamental,many1,many2}. 

Among various classes of duality that have been worked out, 
one particular electric-magnetic pair of duality is 
especially interesting \cite{chi-nonchi,spin10}. 
The electric theory is based on an $Spin(N)$ gauge group with 
matter in vector and spinor representations; 
while the magnetic counterparts have in general an $SU(M)$ 
gauge group with matter in symmetric tensor, fundamental, 
and anti-fundamental representations. 
No rigorous proof on this $Spin-SU$ duality is known, 
but several non-trivial consistency checks support its existence. 
One remarkable feature of the $Spin-SU$ duality appears when 
the spinor representation of $Spin(N)$ is real. 
The electric theory becomes non-chiral. 
However, the matter content of $SU(N)$ magnetic theory is 
obviously in chiral representations, 
the so called chiral$-$non-chiral duality. 

One of the consistency checks on the $Spin-SU$ duality is to 
add masses to all fields in spinor representations. 
The gauge group of the electric theory is broken to 
$SO(N)$.\footnote{ In $Spin(10)$ theory, 
no mass term can be written for spinors. 
Coupling the spinors to a field in vector representation 
that acquires an expectation value can then perform 
the check on the duality. } 
On the magnetic theory, the effect of adding spinor masses 
generally forces the symmetric tensor to acquire expectation 
values and breaks the gauge group $SU(M)$ to 
$SO(M)$.\footnote{ This breaking pattern is characteristic of 
the $Spin(N)$ theory with one spinor representation. } 
After massive fields are integrated out, 
the unbroken $SO(M)$ theory is found to be the dual of 
$SO(N)$ theory under the elementary $SO$ duality of 
\cite{fundamental}. 
This provides strong evidence for the $Spin-SU$ duality.

Recently, Strassler made an interesting observation about 
the breaking pattern of this $Spin-SU$ duality \cite{spinor-monopole}. 
The $SU(M) \to SO(M)$ breaking pattern on the magnetic theory 
renders a non-vanishing second homotopy group 
$\pi_2 [SU(M)/SO(M)]$ = $\pi_1 [SO(M)]$ = $Z_2$, for $M > 2$. 
This indicates that a topological stable monopole carrying a 
$Z_2$ charge exists in the theory \cite{weinberg}. 
He then argued that the $Z_2$ charge associated with the monopole 
of the magnetic theory has to be carried by those massive spinors 
of the electric theory, because the spinors also have a 
$Z_2$ charge under the center of $Spin(N)$, 
i.e. \( Z_2 = Spin(N)/SO(N) \). 
Therefore, the $Z_2$ monopole is the image of massive spinors 
under duality.

It is important to highlight that the monopole in the 
magnetic theory only exists when all spinors are massive. 
If one or more of spinors in the electric theory are massless, 
the massive spinors in general will be surrounded by 
massless spinor cloud and form neutral bound states. 
Under this circumstance, the $Z_2$ charge will not be visible 
in the magnetic theory. 
However, if all spinor fields are massive, the cloud that 
surrounds massive spinors can only be those light fields in 
vector or adjoint representations of $SO(10)$. 
Because these fields are neutral under the group $Z_2$, 
the $Z_2$ charge of massive spinors will not be screened 
at long distance. 
One thus expects that the lightest state with non-zero 
$Z_2$ charge of the electric theory will be visible in the form 
of the heavy monopole in the magnetic theory. 
The strongest evidence of this spinor-monopole identification 
comes from the transformation properties of the two theories 
under flavor groups. 
In \cite{spinor-monopole}, several examples have been given to 
verify that the massive spinors and monopole indeed transform 
in the same way under flavor symmetries. 

In this paper, we continue the work on spinor-monopole 
identification by studying $Spin(10)$ gauge theories with 
matter in $F$ vector and $N$ spinor representations. 
The purpose is to check the transformation properties of 
the monopole under $SU(N)$ flavor symmetry that rotates the 
$N$ spinors into each other. Section 2 briefly describes 
the model for this study. 
The main results of this identification under the $SU(N)$ 
flavor symmetry is contained in Section 3. 
Finally, a conclusion is given in Section 4. 
Note that $SU(F)$ flavor symmetry which rotates $F$ vectors 
into each other will be discussed only peripherally in 
this article because it has been reported in \cite{spinor-monopole}.

\section{The ``expanded" theory of $Spin(10)$ theory}

The electric theory is based on an \( Spin(10)_{\rm local} 
\times [SU(F) \times SU(N) \times U(1)_Y 
\times U(1)_R]_{\rm global} \) symmetry group with 
matter in $F$ vector and $N$ spinor representations. 
The transformation properties of fields under the symmetry group are
\begin{eqnarray}
\label{AField;N}
\PF & \sim & ({\bf 10};{\Box},{\bf 1},-2N,\xi ) \nonumber \\
\Psi & \sim & ({\bf 16};{\bf 1},{\Box},F,0),
\end{eqnarray}
where $\xi = \frac{F+2N-8}{F}$. The superpotential of the 
electric theory is zero, $W = 0$.

The magnetic theory of (\ref{AField;N}) constructed in 
\cite{spin10} is based on an \( SU(F+2N-7) \times Sp(2N-2) \) 
gauge group. 
One remarkable structure worthy of attention is the partial 
$SU(2)$ realization of the $SU(N)$ flavor symmetry 
in the magnetic theory. 
The full $SU(N)$ flavor symmetry only exist as a quantum 
accidental symmetry \cite{accidental}. 
However, when $N=2$, the flavor symmetries of the electric theory 
and the magnetic theory are equivalent. 
(The magnetic theory is summarized in appendix B). 

If one wishes to analyze the problem of spinor-monopole 
identification on the magnetic theory, he/she will soon discover 
that the complex and intricate breaking pattern in the theory 
is too difficult to perform. We thus use a dual description 
of the magnetic theory for study. 
This new description is derived from expanding the symmetric tensor 
of the magnetic theory with the method of deconfinement. 
According to \cite{decon1,decon2}, there are two such approaches 
available. The first approach treats the symmetric tensor as 
a bound state of an auxiliary gauge theory, whereas the second 
approach considers the symmetric tensor and the fundamentals as 
bound states of another auxiliary gauge theory. 
Both approaches are able to yield dual descriptions of the 
magnetic theory. Yet, the latter method suggested in \cite{decon2} 
renders a good theory with controllable breaking pattern when 
masses of spinors are introduced. 

This new theory is based on an  \( [SO(10) \times SU(2N+4) \times 
Sp(2N-2)]_{\rm local} \times [SU(F) \times SU(2) \times U(1)_Y 
\times U(1)_R]_{\rm global} \) symmetry group and will be called 
the ``expanded" theory of the $Spin(10)$ theory. 
As in the magnetic theory, the $SU(N)$ flavor symmetry is also 
partially realized in this theory. 
The field content is
\begin{eqnarray}
\label{CField;N}
\PF & \sim & ({\Box},{\bf 1},{\bf 1};
{\Box},{\bf 1},-2N,\xi) 
\nonumber  \\ 
S & \sim & ({\bf 1},{\iiBox},{\bf 1};
{\bf 1},{\bf 1},2F-\frac{6F}{N+2},\frac{1}{N+2}) 
\nonumber  \\ 
Q & \sim & ({\bf 1},{\Box},{\bf 1};
{\bf 1},{\bf 2N-1},-F-\frac{3F}{N+2},1+\frac{1}{2(N+2)}) 
\nonumber  \\ 
\bQP & \sim & ({\Box},{\bBox},{\bf 1};
{\bf 1},{\bf 1},-F+\frac{3F}{N+2},1-\frac{1}{2(N+2)}) 
\nonumber  \\ 
\bQN & \sim & ({\bf 1},{\bBox},{\bf 1};
{\bf 1},{\bf 1},(2N-1)F+\frac{3F}{N+2},-2(N+1)-\frac{1}{2(N+2)}) 
\\ 
\bQ & \sim & ({\bf 1},{\bBox},{\Box};
{\bf 1},{\bf 2},-F+\frac{3F}{N+2},1-\frac{1}{2(N+2)}) 
\nonumber  \\ 
R & \sim & ({\bf 1},{\bf 1},{\Box};{\bf 1},{\bf 2N-2},2F,0) 
\nonumber  \\
P & \sim & ({\Box},{\bf 1},{\bf 1};{\bf 1},{\bf 2N-1},2F,0) 
\nonumber \\
N & \sim & ({\bf 1},{\bf 1},{\bf 1};{\bf 1},{\bf 2N-1},-2(N-1)F,2N+3) 
\nonumber  \\
u & \sim & ({\bf 1},{\bf 1},{\bf 1};{\bf 1},{\bf 1},-4NF,4N+6). 
\nonumber
\end{eqnarray}
The ``expanded" theory has this superpotential
\begin{equation}
W = \bQP^2 S + u \bQN^2 S + \bQ^2 S + P \bQP Q + N \bQN Q +
      \bQ R Q.
\label{Csuperpotential;N}
\end{equation}
Note that all parameters, including those needed for dimensional 
consistency, are set to unity.

At first glance, the ``expanded" theory is far more complicated 
than the magnetic theory because the former is consisted of a 
product of three gauge groups and many chiral superfields. 
Yet, it will be shown later that this poses no obstacle to 
calculation. 
As can be seen from the matter content, all fields in the 
``expanded" theory, except for $P_F$, transform as singlet fields 
under $SU(F)$ flavor symmetry. 
The ``expanded" theory is thus like the $Spin(10)$ theory with 
spinor representations being substituted by vector representations 
of an $SO(10)$ theory. 
This $SO(10)$ theory is then extended to a theory with three 
gauge groups for the purposes that all local anomalies will cancel, 
all global anomalies can match, and all composite operators can 
map to those of the $Spin(10)$ theory. 

The term ``expanded" is derived from the fact that a product of 
two $Spin(N)$ spinors can always be decomposed into a direct sum of 
anti-symmetric tensors of $SO(N)$. 
For instance, the product of two $Spin(10)$ spinors has this 
decomposition: \( {\bf 16} \times {\bf 16} = [{\bf 1}]_{\rm S} + 
[{\bf 3}]_{\rm A} + [\tilde{\bf 5}]_{\rm S} \). 
Here, $[{\bf n}]$ denotes an anti-symmetric rank-n tensor 
representation, ``S" and ``A" subscripts indicate symmetry and 
anti-symmetry under spinor exchange, and the tilde over the last 
term implies that the rank-5 representation is complex self-dual. 
These anti-symmetric tensors are then further deconfined into 
the fields in fundamental representations of the 
\( SO(10) \times SU(2N+4) \times Sp(2N-2) \) gauge theory. 
That is,\footnote{ In principle, the ``expanded" theory of any 
$Spin(N)$ theory can be determined from this rule. }
\begin{eqnarray}
\Psi^2_{[{\bf 1}]} & \longleftrightarrow & P + \bQP \bQ^{2N+2} \bQN   
\nonumber \\
\Psi^2_{[{\bf 3}]} & \longleftrightarrow & \bQP^3 \bQ^{2N} \bQN 
\\
\Psi^2_{[{\bf 5}]} & \longleftrightarrow & \bQP^5 \bQ^{2N-2} \bQN,
\nonumber
\end{eqnarray}   
\label{Cmapping;N}
where the subscript $[{\bf n}]$ denotes an $SO(10)$ anti-symmetric 
tensor of rank-$n$.

Does the ``expanded" theory have a solution of topological stable 
monopole as the magnetic theory has when masses of all spinors 
are introduced? 
The answer is yes. 
Because all fields in the ``expanded" theory are neutral under 
the $Z_2$ group, they cannot screen the topological charge. 
The $Z_2$ charge has to be carried by a monopole of the 
``expanded" theory. 
This statement can be easily checked for one spinor case, $N=1$. 
On one side, the $Spin(10)$ theory is broken to $SO(9)$ upon 
introducing a coupling of spinor to a vector that an acquires 
expectation value. 
On the other side, the $SO(10) \times SU(6)$ gauge group of the 
``expanded" theory is first broken to $SO(9) \times SU(5)$ and then 
to $SO(9) \times SO(5)$. 
At the latter stage of breaking sequences, a $Z_2$ monopole will 
be generated since \(  \pi_2 [SU(5)/SO(5)] = Z_2 \). 
($N=1$ $Spin(10)$ theory and its ``expanded" theory are summarized 
in appendix A). 
We will show in next section that the breaking pattern 
\( SU(5) \to SO(5) \) also exists for $N \ge 2$ spinor cases. 
A remark is noted. The $SO(9) \times SO(5)$ ``expanded" theory will 
be reduced to an $SO(9)$ theory since the $SO(5)$ theory confines. 
After massive fields are integrated out, this $SO(9)$ theory is 
equivalent to the electric $SO(9)$ theory. The ``expanded" theory 
is thus self-dual, rather than dual, to the $Spin(10)$ theory 
under duality.

\section{The spinor-monopole identification}

In this section, the ``expanded" theory (\ref{CField;N}) is chosen 
for the study of spinor-monopole identification of the $Spin(10)$ 
theory with matter in $F$ vector and $N$ spinor representations. 
Because the spinor representation of $Spin(10)$ is chiral, 
no mass can be introduced for spinor fields. 
However, mass terms can be given when $Spin(10)$ is broken to a 
smaller $Spin$ group. 
In the $Spin(10)$ theory (\ref{AField;N}), by coupling spinors to 
a vector that acquires an expectation value, the spinors become 
massive in an $Spin(9)$ theory. 
This coupling also breaks the ``expanded" theory (\ref{CField;N}) 
to an \( [SO(9) \times SU(2N+4) \times Sp(2N-2)]_{\rm local} 
\times [SU(F-1) \times SU(2)]_{\rm global} \) theory, which differs 
from (\ref{CField;N}) with regard to $SO$ gauge group. 
In other words, this new theory will have matter content as in 
(\ref{CField;N}) except for those fields in vector representations 
of $SO(10)$. 
The $SO(10)$ vectors $\bQP$ and $P$ of (\ref{CField;N}) get split 
according to this: \( \bQP \to (\bQP,\bqP) \) and \( P \to (P,p) \), 
where $\bQP$ and $P$ are $SO(9)$ vectors and $\bqP$ and $p$ are 
$SO(9)$ scalars. 
Note that we use the same notation for both $SO(10)$ and $SO(9)$ 
vectors. 
Taking the effects of field splitting and mass perturbation into 
accounts, we conclude that this theory has the effective superpotential
\begin{eqnarray}
W_{N,j} & = & \sum_{a,b=1}^{2N+4} \Biggl\{
        S^{ab} \bQP^a \bQP^b + u S^{ab} \bQN^a \bQN^b + 
        S^{ab} \bqP^a \bqP^b +
        \sum_{i=1}^{2N-1} \biggl[  P_i Q_i^a \bQP^a + 
        N_i Q_i^a \bQN^a + p_i Q_i^a \bqP^a \biggr]  \nonumber \\
  &  & + \sum_{\alpha ,\beta =1}^{2N-2} J_{\alpha \beta} \biggl[
       S^{ab} \big( \bQ_1^{a,\alpha} \bQ_2^{b,\beta} -
              \bQ_2^{a,\alpha} \bQ_1^{b,\beta} \big) +
       \sum_{i=1}^{2N-1} Q_i^a 
              \big( \bQ_1^{a,\alpha} R_i^\beta + 
              \bQ_2^{a,\alpha} R_{i-1}^\beta \big) 
       \biggr] \Biggr\} + p_j \, .
\label{W9;N}
\end{eqnarray}
where the notation $W_{N,j}$ denotes that the superpotential of an 
\( SO(9) \times SU(2N+4) \times Sp(2N-2) \) theory is perturbed by 
$p_j$ in the $j$-th direction. 
Here, $a,b$ superscripts denote $SU(2N+4)$ gauge indices, 
$\alpha, \beta$ superscripts indicate $Sp(2N-2)$ gauge indices, 
and $i,j$ subscripts are $SU(2)$ flavor indices. 
$J_{\alpha \beta}$ is an anti-symmetric tensor taken to be 
\( J = {\bf 1}_{N-1} \otimes i \sigma_2 \). 
\( R_0^\alpha = R_{2N-1}^\alpha = 0 \) are introduced to make 
the expression compact. All Clebsch-Gordan coefficients of 
$SU(2)$ multiplication are set to unity for simplicity. 
The mass term introduced for spinors in the electric theory is 
mapped to the last term $p_j$ in (\ref{W9;N}). 
The other candidate $\bqP \bQ^{2N+2} \bQN$ that also maps to the 
mass term is of no interest to the dynamics, 
and will not be discussed. 
It is noted that masses of all spinors are taken to be equal, 
and $SO(9)$ gauge indices are omitted because they are spectators 
to the breaking pattern of Higgs mechanism.    

In (\ref{W9;N}), the F-flatness condition $\p W/\p p_j$ ensures 
that $\langle \bqP^a Q^a_j \rangle$ is nonzero, and the gauge group 
of the theory will be broken to a smaller group. 
The detailed breaking pattern is of course dependent on 
what direction of $p_j$ in the $(2N-1)$ representation of $SU(2)$ 
is assigned to. 
Because of the reflection symmetry of $SU(2)$, there are only $N$ 
possibilities for $p_j$. 
It will be shown below that the choice of \( p_j = p_N \) in 
(\ref{W9;N}) corresponds to grant masses to all spinors in 
the electric theory. 
A monopole state which carries a $Z_2$ topological charge is found 
to exist in the ``expanded" theory at the final stage of 
Higgs breaking sequences. 
According to the prediction of \cite{spinor-monopole}, this monopole 
will be accompanied by $2(N-1)$ zero modes that after quantization 
make the monopole an $(N-1)$-index multispinor under the 
$SU(2)$ flavor symmetry.\footnote{ 
There are other types of zero modes that make the same monopole an 
multispinor under another flavor symmetry. } 
Next, this prediction will be checked on the ``expanded" theory.

\subsection{The $Spin(10)$ theory with two spinors}

For two-spinor case, $N=2$, the ``expanded" theory is based on an 
\( SU(8) \times Sp(2) \) gauge group. 
(The notation of $SO(9)$ gauge group is suppressed). 
The scalar field $p_j$ forms a three dimensional representation 
of $SU(2)$, that is, \( p_j = (p_1,p_2,p_3) \). 

First, let us investigate the choice of $p_j = p_1$ in (\ref{W9;N}). 
The F-flatness condition of $p_1$ implies that $\bqP^a$ and $Q^a_1$ 
get expectation values, which by the D-term equation must be 
in the same direction of the color group. 
Thus, the gauge group is broken to \( SU(7) \times Sp(2) \) with 
the superpotential 
\begin{eqnarray}
W_{2,1}^{(1)} 
& = & \sum_{a,b=1}^{7} \Biggl\{
        S^{ab} \bQP^a \bQP^b + u S^{ab} \bQN^a \bQN^b + 
        \sum_{i=2}^{3} \biggl[  P_i Q_i^a \bQP^a + 
        N_i Q_i^a \bQN^a \biggr]  
\nonumber \\
&   & + \sum_{\alpha ,\beta =1}^{2} J_{\alpha \beta} \biggl[
       Q_2^a \bQ_1^{a,\alpha} R_2^\beta + 
       Q_3^a \bQ_2^{a,\alpha} R_2^\beta +
       S^{ab} \bigl( \bQ_1^{a,\alpha} \bQ_2^{b,\beta} -
              \bQ_2^{a,\alpha} \bQ_1^{b,\beta} \bigr) 
\nonumber \\
&   & + \,
       S^{8a} \bigl( Q_2^b \bQ_2^{b,\alpha} \bQ_2^{a,\beta} -
                           \bQ_2^{8,\alpha} \bQ_1^{a,\beta} \bigr) +
       S^{88} Q_2^a \bQ_2^{a,\alpha} \bQ_2^{8,\beta}  
       \biggr] \Biggr\} + S^{88},
\label{W9;21;1}
\end{eqnarray}
where we pick \( \langle \bqP^8 \rangle \langle Q_1^8 \rangle \ne 0 \) 
and the superscript $(n)$ denotes that the superpotential 
$W_{2,1}^{(n)}$ is derived after $n$ steps of symmetry breaking.  

Now work out the F-flatness conditions for this superpotential. 
The vanishing \( \p W_{2,1}^{(1)}/\p S^{88} \) condition ensures 
that \( \langle J_{\alpha \beta} Q_2^a \bQ_2^{a,\alpha} 
\bQ_2^{8,\beta} \rangle \) is non-zero. 
\( \langle Q_2^7 \rangle \langle \bQ_2^{7,1} \rangle \langle 
\bQ_2^{8,2} \rangle \ne 0 \) is adopted in computation. 
This breaks the $SU(7)$ gauge symmetry down to $SU(6)$ and 
completely breaks the $Sp(2)$ gauge group in which a dynamical 
superpotential is generated by a weak coupling instanton process. 
After combining this quantum correction with the classical terms 
(\ref{W9;21;1}), we obtain the superpotential of this 
$SU(6)$ gauge theory,
\begin{equation}
W_{2,1}^{(2)}  = 
\sum_{a,b=1}^{6} \Biggl\{
     S^{ab} \bQP^a \bQP^b + u S^{ab} \bQN^a \bQN^b + 
     P_3 Q_3^a \bQP^a + N_3 Q_3^a \bQN^a +
     S^{ab} \bq^a \bq^a +
     R_3^1 Q_3^a \bq^a + \det S^{ab}  \Biggr\},
\label{W9;21;2}
\end{equation}
where \( \bq^a \equiv \bQ_1^{a,1} = \bQ_2^{a,2} \) and the last term 
on the equation comes from the instanton effect. 

Note that the theory with superpotential (\ref{W9;21;2}) is the 
``expanded" theory of an $Spin(9)$ theory with matter in $(F-1)$ 
vector and one spinor representations. 
Readers can confirm this point from appendix A. 
Hence, the choice of $p_j = p_1$ in the \( SU(8) \times Sp(2) \) 
``expanded" theory is mapped to a mass perturbation for one of 
the two spinors in the $Spin(9)$ theory with $N=2$. 
As mentioned earlier, a similar breaking result, 
\( SU(8) \times Sp(2) \to SU(7) \times Sp(2) \to SU(6) \), is 
obtainable when \( p_j = p_3 \). 

Next, let us consider the same \( SU(8) \times Sp(2) \) gauge theory 
with this choice $p_j = p_2$. 
The theory is first broken to an \( SU(7) \times Sp(2) \) theory 
by $\langle \bqP^a Q_2^a \rangle \equiv \langle \bqP^8 \rangle 
\langle Q_2^8 \rangle \ne 0$ with the superpotential 
\begin{eqnarray}
W_{2,2}^{(1)} 
& = & \sum_{a,b=1}^{7} \Biggl\{
      S^{ab} \bQP^a \bQP^b + u S^{ab} \bQN^a \bQN^b + 
      \sum_{i=1,3} \biggl[
      P_i Q_i^a \bQP^a + N_i Q_i^a \bQN^a \biggr] 
\nonumber \\
&   & + \,
      \sum_{\alpha ,\beta =1}^{2} J_{\alpha \beta} \biggl[
      S^{ab} \bigl( \bQ_1^{a,\alpha} \bQ_2^{b,\beta} -
             \bQ_2^{a,\alpha} \bQ_1^{b,\beta} \bigr) +
      S^{8a} \bigl( Q_3^b \bQ_2^{b,\alpha} \bQ_2^{a,\beta} -
                    Q_1^b \bQ_1^{b,\alpha} \bQ_1^{a,\beta} \bigr)
\nonumber \\
&   & + \,
      S^{88} Q_1^a \bQ_1^{a,\alpha} Q_3^b \bQ_2^{b,\beta}
      \biggr] \Biggr\} + S^{88}.
\label{W9;22;1}
\end{eqnarray}

In a similar vein, by working on the F-flatness condition 
$\p W_{2,2}^{(1)}/\p S^{88}$ of this superpotential, 
we arrive at a non-vanishing 
\( \langle J_{\alpha \beta} Q_1^a \bQ_1^{a,\alpha} 
Q_3^b \bQ_2^{b,\beta} \rangle \). 
This indicates that the $SU(7) \times Sp(2)$ gauge group is further 
broken to $SU(5)$ in addition to a non-perturbative superpotential 
that is generated by $Sp(2)$ instanton process. 
The result of the choice of expectation values, 
\( \langle Q_1^7 \rangle \langle \bQ_1^{7,1} \rangle 
\langle Q_3^6 \rangle \langle \bQ_2^{6,2} \rangle \), 
produces this superpotential
\begin{equation}
W_{2,2}^{(2)}  = 
\sum_{a,b=1}^{5} \Biggl\{
     S^{ab} \bQP^a \bQP^b + u S^{ab} \bQN^a \bQN^b + 
     S^{ab} \bq^a \bq^a +
     S^{6a} S^{7b} [S^4]^{ab} + S^{67} \det S^{ab} \Biggr\} + S^{67},
\label{W9;22;2}
\end{equation}
where \( [S^4]^{ab} \equiv \epsilon^{aa_1 \cdots a_4} 
\epsilon^{bb_1 \cdots b_4} S^{a_1b_1} \cdots S^{a_4b_4} \) and 
the last two terms inside the brace signs are from 
instanton contribution. 
Another consistency checks of (\ref{W9;22;2}) is provided in appendix C. 

Yet, the symmetry breaking does not cease at theory (\ref{W9;22;2}). 
The vanishing of the F-flatness condition for field $S^{67}$ 
ensures that \( \langle \det S^{ab} \rangle \) is nonzero. 
This will break the gauge group $SU(5)$ down to $SO(5)$ and thus 
generates a $Z_2$ monopole solution. 
Before massive fields are integrated out, the $SO(5)$ gauge theory 
has the superpotential
\begin{equation}
W_{2,2}^{(3)}  = 
\sum_{a=1}^{5}  \Biggl\{   
\langle S \rangle \bQP^a \bQP^a + \langle S \rangle u \bQN^a \bQN^a + 
\langle S \rangle \bq^a \bq^a +
\langle S^4 \rangle S^{6a} S^{7a} \Biggr\}.
\label{W9;22;3}
\end{equation}

As can be seen from (\ref{W9;22;3}), fields $\bQP^a$, $\bq^a$, $S^{6a}$,
and $S^{7a}$ become massive as a result of their coupling to $S$. 
In the presence of the monopole, $\bQP^a$ bears an extra $SO(9)$ 
gauge index and has nine zero modes, making the monopole an 
$Spin(9)$ spinor after quantization. 
Similarly, $\bq^a$ has one zero mode and will make the monopole an 
``$Spin(1)$" spinor. 
Analogously, $S^{6a}$ and $S^{7a}$ each has a single zero mode that, 
upon quantization, constructs the monopole a two dimensional 
spinor under $SU(2)$ flavor symmetry. 
This agrees with the transformation properties of the spinors of 
the electric theory under the $Spin(9)$ color group and 
the $SU(2)$ flavor group. 
We thus verify the prediction of spinor-monopole identification 
in the model of $Spin(10)$ gauge theory with two spinors.

As a side remark, after all massive fields are integrated out, 
the $SO(5)$ theory confines without generating a superpotential. 
The final result is a pure $SO(9)$ theory with matter in $(F-1)$ 
vector representations and a heavy monopole. 
Under duality, this theory is self-due to the $Spin(9)$ gauge theory 
with $(F-1)$ vectors and two massive spinors.

\subsection{The $Spin(10)$ theory with $N \ge 3$ spinors}

For $N$-spinor case, the ``expanded" theory (\ref{W9;N}) has an 
\( SU(2N+4) \times Sp(2N-2)\) gauge group. 
The field $p_j$ forms a $(2N-1)$ dimensional representation of 
$SU(2)$, that is, \( p_j = (p_1,\cdots,p_{2N-1}) \). 
The reflection symmetry of $SU(2)$ connotes that the choices of 
$p_k$ and $p_{2N-k}$ \( (1 \le k \le N) \) generate the same symmetry
breaking pattern. 
We thus focus on the range of $p_j$ with \( j=1, \cdots , N \).

As discussed in Section 3.1, it can be demonstrated that, 
for the choice of $p_j$ with $j \le N-2$, 
the theory is eventually broken to an 
\( SU(2(N-j)+4) \times Sp(2(N-j)-2) \) gauge symmetry in which 
the superpotential takes the form $W_{N-j,0}$ (see (\ref{W9;N})). 
Here the subscript ``0" indicates the removal of the field $p_j$ 
in (\ref{W9;N}). 
(The result is summarized in appendix D.) 
This theory is apparently the ``expanded" theory of an $Spin(9)$ 
gauge theory with matter in $(F-1)$ vector and $(N-j)$ 
spinor representations. 
Similarly, for \( p_j = p_{N-1} \) situation, the resulting theory 
is based on an $SU(6)$ gauge symmetry, the ``expanded" theory of 
an $Spin(9)$ theory with $(F-1)$ vectors and one spinor.

Let us continue by choosing \( j=N \) for $p_j$. The final theory 
will be reduced to an $SO(9)$ gauge theory with matter in 
$(F-1)$ vector representations and a monopole solution. 
This can be illustrated as follows. 
The F-flatness condition \( \p W_{N,N} / \p p_N \) of (\ref{W9;N}) 
implies that $ \langle \bqP^a Q^a_N \rangle$ is non-zero. 
Thus the \( SU(2N+4) \times Sp(2N-2) \) gauge group is broken to 
\( SO(2N+3) \times Sp(2N-2) \) with superpotential generically 
denoted by $W_{N,N}^{(1)}$. 
Following this superpotential would be the second step of symmetry 
breaking, it produces a theory based on an 
\( SO(2N+1) \times Sp(2N-4) \) gauge group with superpotential 
\begin{eqnarray}
W_{N,N}^{(2)} 
& = & \sum_{a,b=1}^{2N+1} \Biggl\{
      S^{ab} \bQP^a \bQP^b + u S^{ab} \bQN^a \bQN^b + 
      S^{ab} \bq^a \bq^b +
      \mathop{{\sum}'}_{i=1}^{2N-1} 
      \biggl[ P_i Q_i^a \bQP^a + N_i Q_i^a \bQN^a + 
      h_i Q_i^a \bq^a  \biggr]  
\nonumber \\
&   & + \sum_{\alpha ,\beta =1}^{2N-4} J_{\alpha \beta} \biggl[
      S^{ab} \bigl( \bQ_1^{a,\alpha} \bQ_2^{b,\beta} -
             \bQ_2^{a,\alpha} \bQ_1^{b,\beta} \bigr) +
      \mathop{{\sum}'}_{i=1}^{2N-1}
      Q_i^a \big( \bQ_1^{a,\alpha} R_i^\beta + 
            \bQ_2^{a,\alpha} R_{i-1}^\beta \big)
\nonumber \\
&   & + \,
      S^{2N+2,a} Q_{N+2}^b \bQ_2^{b,\alpha} \bQ_2^{a,\beta} +
      S^{2N+3,a} Q_{N-2}^b \bQ_1^{b,\alpha} \bQ_1^{a,\beta} + 
      Q_1^{2N+2} Q_{N+2}^a \bQ_2^{a,\alpha} R_1^\beta      
\nonumber \\
&   & + \,
      Q_{2N-1}^{2N+3} Q_{N-2}^a \bQ_1^{a,\alpha} R_{2N-2}^\beta +
      S^{2N+2,2N+3} Q_{N-2}^a \bQ_1^{a,\alpha} Q_{N+2}^b \bQ_2^{b,\beta}
      \biggr] \Biggr\} + S^{2N+2,2N+3},
\label{W9;NN;2}
\end{eqnarray}
where \( \mathop{{\sum}'}_{i=1}^{2N-1} \) denotes that the sum is 
taken from the interval \( (1,\cdots,N-2,N+2,\cdots,2N-1) \). 
The fields \( \bq^a \equiv \bQ_1^{a,2N-2} = \bQ_2^{a,2N-3} \), 
\( h_i \equiv (R_i^{2N-2} + R_{i-1}^{2N-3}) \), and 
\( R_0^\alpha = R_{N-2}^\alpha = R_{N+1}^\alpha = 
R_{2N-1}^\alpha = 0 \) are introduced to pack the expression.

The theory (\ref{W9;NN;2}) is further broken to a smaller gauge 
group by the F-flatness condition of $S^{2N+2,2N+3}$. 
The result of integrating out massive fields is 
an \( SU(2N-1) \times Sp(2N-6) \) gauge group with the superpotential
\begin{eqnarray}
W_{N,N}^{(3)} 
& = & \sum_{a,b=1}^{2N-1} \Biggl\{
      S^{ab} \bQP^a \bQP^b + u S^{ab} \bQN^a \bQN^b + 
      S^{ab} \bq^a \bq^b +
      \mathop{{\sum}''}_{i=1}^{2N-1} 
      \biggl[ P_i Q_i^a \bQP^a + N_i Q_i^a \bQN^a + 
      h_i Q_i^a \bq^a  \biggr]  
\nonumber \\
&   & + \sum_{\alpha ,\beta =1}^{2N-6} J_{\alpha \beta} \biggl[
      S^{ab} \bigl( \bQ_1^{a,\alpha} \bQ_2^{b,\beta} -
             \bQ_2^{a,\alpha} \bQ_1^{b,\beta} \bigr) +
      \mathop{{\sum}''}_{i=1}^{2N-1}
      Q_i^a \big( \bQ_1^{a,\alpha} R_i^\beta + 
            \bQ_2^{a,\alpha} R_{i-1}^\beta \big)
\nonumber \\
&   & + \,
      S^{2N,a} Q_{N+3}^b \bQ_2^{b,\alpha} \bQ_2^{a,\beta} +
      S^{2N+1,a} Q_{N-3}^b \bQ_1^{b,\alpha} \bQ_1^{a,\beta} + 
      Q_1^{2N} Q_{N+3}^a \bQ_2^{a,\alpha} R_1^\beta      
\nonumber \\
&   & + \,
      Q_{2N-1}^{2N+1} Q_{N-3}^a \bQ_1^{a,\alpha} R_{2N-2}^\beta +
      S^{2N,2N+1} Q_{N-3}^a \bQ_1^{a,\alpha} Q_{N+3}^b \bQ_2^{b,\beta}
      \biggr] \Biggr\} + S^{2N,2N+1},
\label{W9;NN;3}
\end{eqnarray}
where \( \mathop{{\sum}''}_{i=1}^{2N-1} \) denotes that the sum is 
taken from \(1,\cdots,N-3,N+3,\cdots,2N-1 \) and 
\( R_0^\alpha = R_{N-3}^\alpha = R_{N+2}^\alpha = R_{2N-1}^\alpha = 0 \).

Note that superpotential (\ref{W9;NN;3}) is similar to superpotential 
(\ref{W9;NN;2}) except that the former has less numbers of superfields. 
The two superpotentials differ only in the summation over 
gauge indices \( a,b \) and \( \alpha ,\beta \) and over 
the $SU(2)$ flavor index $i$. 
The similarity of the two superpotentials (\ref{W9;NN;2}) 
and (\ref{W9;NN;3}) suggests that the structure of superpotential 
is quite unique and can be formally expressed in the same form 
at each stage of symmetry breaking sequences. 
Hence, we would expect that the superpotential (\ref{W9;NN;3}) 
will be reduced to an superpotential, denoted by $W_{N,N}^{(4)}$, 
of the same form. 
In a similar way, $W_{N,N}^{(4)}$ will be reduced to $W_{N,N}^{(5)}$ 
of the same form, $W_{N,N}^{(5)}$ will be reduced to $W_{N,N}^{(6)}$ 
of the same form, and this pattern will continue to hold true 
before the $Sp$ gauge group is completely broken. 

This finding is significant, because it simplifies huge amounts of 
analysis on the breaking pattern. 
It assists us in constructing the breaking pattern of the 
``expanded" theory in a systematic way. 
That is, when an \( SU(2N+4) \times Sp(2N-2) \) ``expanded" theory 
(\ref{W9;N}) is perturbed by $p_j$ along the $N$-th direction, 
the breaking pattern has this sequence: \( 
SU(2N+4) \times Sp(2N-2) \stackrel{(1)}{\longrightarrow} 
SU(2N+3) \times Sp(2N-2) \stackrel{(2)}{\longrightarrow} 
SU(2N+1) \times Sp(2N-4) \stackrel{(3)}{\longrightarrow} 
\cdots \stackrel{(N-2)}{\longrightarrow}
SU(9) \times Sp(4) \stackrel{(N-1)}{\longrightarrow} 
SU(7) \times Sp(2) \), with the superpotential takes the same form 
at each breaking stage. 
Therefore, the superpotential of the above \( SU(7) \times Sp(2) \) 
theory will have this form
\begin{eqnarray}
W_{N,N}^{(N-1)} 
& = & \sum_{a,b=1}^{7} \Biggl\{
      S^{ab} \bQP^a \bQP^b + u S^{ab} \bQN^a \bQN^b +
      S^{ab} \bq^a \bq^b +
      \sum_{i=1,2N-1} \biggl[
      P_i Q_i^a \bQP^a + N_i Q_i^a \bQN^a + 
      h_i Q_i^a \bq^a \biggr] 
\nonumber \\
&   & + \,
      \sum_{\alpha ,\beta =1}^{2} J_{\alpha \beta} \biggl[
      S^{ab} \bigl( \bQ_1^{a,\alpha} \bQ_2^{b,\beta} -
             \bQ_2^{a,\alpha} \bQ_1^{b,\beta} \bigr) +
      S^{8a} Q_{2N-1}^b \bQ_2^{b,\alpha} \bQ_2^{a,\beta} +
      S^{9a} Q_1^b \bQ_1^{b,\alpha} \bQ_1^{a,\beta} 
\nonumber \\
&   & + \,
      S^{89} Q_1^a \bQ_1^{a,\alpha} Q_{2N-1}^b \bQ_2^{b,\beta}
      \biggr] \Biggr\} + S^{89}.
\label{W9;NN;N-1}
\end{eqnarray}
Interested readers may compare and contrast 
(\ref{W9;22;1}) and (\ref{W9;NN;N-1}).  

Now, the F-flatness condition $\p W_{N,N}^{(N-1)}/\p S^{89}$ of 
this superpotential implies that the $SU(7) \times Sp(2)$ gauge group 
is broken to $SU(5)$ with a non-perturbative superpotential 
generated by $Sp(2)$ instanton effects. 
This $SU(5)$ theory has superpotential
\begin{equation}
W_{N,N}^{(N)}  = 
\sum_{a,b=1}^{5} \Biggl\{
     S^{ab} \bQP^a \bQP^b + u S^{ab} \bQN^a \bQN^b + 
     S^{ab} \bq^a \bq^a +
     S^{6a} S^{7b} [S^4]^{ab} + S^{67} \det S^{ab} \Biggr\} + S^{67}.
\label{W9;NN;N}
\end{equation}

By comparing (\ref{W9;NN;N}) and (\ref{W9;22;2}), we find that they 
are equivalent. 
As a result, the discussion on the spinor-monopole identification of 
(\ref{W9;22;2}) in previous section can be applied directly to 
(\ref{W9;NN;N}) without any modification. 
That is, the $SU(5)$ gauge symmetry will be broken to $SO(5)$ with 
superpotential \( W_{N,N}^{(N+1)} = W_{2,2}^{(3)} \) and generates 
a $Z_2$ monopole solution. 
Under $SU(2)$ flavor group, the monopole is accompanied with two 
zero modes which make it a two dimensional spinor. 
Unfortunately, this goes against the prediction of 
\cite{spinor-monopole} that suggests the existence of $2(N-1)$ 
zero modes for \( N \ge 3 \) theories.

\section{Conclusion}

The ``expanded" description presented in this article for $Spin(10)$ 
gauge theories with matter in $F$ vector and $N$ spinor representations 
is used for the study of spinor-monopole identification proposed by 
Strassler. 
Because this theory is derived by deconfining the magnetic theory, 
the monopole solution is shown to exist in the ``expanded" theory as 
it does in the magnetic theory. 
This ``expanded" description has a systematic breaking pattern of 
gauge symmetry, upon all spinors become massive in 
the $Spin(10)$ theory.

For the $Spin(10)$ theory with two spinors, we confirm this 
identification by matching the transformation properties of the 
two theories under $SU(2)$ flavor symmetry. 
We find that the existence of two zero modes in the ``expanded" 
theory makes the monopole an $SU(2)$ spinor that corresponds to 
two massive spinors of $Spin(9)$ theory. 
We also note that the monopole transforms as an $Spin(9)$ spinor 
as the massive spinors do. 
However, for theories with $N \ge 3$ spinor fields, the transformation 
properties are not matched between the massive spinors and the monopole. 

This discrepancy for $N \ge 3$ may be due to the partial $SU(2)$ 
realization of the $SU(N)$ flavor symmetry in the ``expanded" theory 
(and in the magnetic theory). 
It can be inferred by breaking the $Spin(10)$ theory down to 
$Spin(10-k)$ with expectation values of $k$ vectors, then the 
electric theory has an $SU(N) \times Spin(k)$ flavor symmetry, 
with all spinors massive.\footnote{ 
The $SU(F-k)$ flavor symmetry is suppressed. } 
On the other side, the ``expanded" theory would have an 
\( [SO(10-k) \times SO(5)]_{\rm local} \times 
[SU(2) \times SO(k)]_{\rm global} \) symmetry group. 
It can be demonstrated that the transformation properties of 
the spinors and the monopole are identical under the $Spin(k)$ 
flavor symmetry, but not so under the $SU(2)$ flavor symmetry. 
We thus speculate that the partial realization of flavor symmetry 
is responsible for the inconsistency. 
However, to resolve this needs further investigation on the 
quantum accident symmetry \cite{accidental}.


The author is grateful to W.-L. Lin for useful discussions. 
This work was supported by NSC grant \# 86-2112-M009-034T, 
Center for Theoretical Sciences of NSC of R.O.C., and 
a grant from  National Tsing-Hua university.

\appendix

\section{appendix}

This appendix summarizes the $Spin(10)$ theory with matter in $F$ 
vector and one spinor representations. 
For the purpose of simplicity, we set all coupling constants and 
dimensional parameters to unity. 
For details, please refer to \cite{chi-nonchi}.

The electric theory is based on an $Spin(10)$ gauge group. 
Under the symmetry groups \( Spin(10)_{\rm local} \times 
[SU(F) \times U(1)_Y \times U(1)_R]_{\rm global} \), the fields of 
the theory transform as follows,
\begin{eqnarray}
\label{AFields;1}
\PF & \sim & ({\bf 10};{\Box},-2,1-\frac{6}{F})  \nonumber \\
\Psi & \sim & ({\bf 16}; {\bf 1}, F,0). 
\end{eqnarray}
The superpotential of the electric theory is zero, $W = 0$.

The ``expanded" theory has an $SO(10) \times SU(6)$ gauge group. 
Under the symmetry group, \( [SO(10) \times SU(6)]_{\rm local} \times 
[SU(F) \times U(1)_Y \times U(1)_R]_{\rm global} \), 
the matter content is
\begin{eqnarray}
\label{CField;1}
\PF & \sim & ({\Box},{\bf 1};{\Box},-2,1-\frac{6}{F})  
\nonumber \\
S & \sim & ({\bf 1},{\iiBox};{\bf 1},0,\frac{1}{3}) 
\nonumber  \\
Q & \sim & ({\bf 1},{\Box};{\bf 1},-2F,\frac{7}{6})  
\nonumber \\
\bQP & \sim & ({\Box},{\bBox};{\bf 1},0,\frac{5}{6})  
\\
\bQN & \sim & ({\bf 1},{\bBox};{\bf 1},2F,-\frac{25}{6})  
\nonumber \\
P & \sim & ({\Box},{\bf 1};{\bf 1},2F,0)  
\nonumber \\
N & \sim & ({\bf 1},{\bf 1};{\bf 1},0,5)
\nonumber \\
u & \sim & ({\bf 1},{\bf 1};{\bf 1},-4F,10). 
\nonumber
\end{eqnarray}
The superpotential of the ``expanded" theory is
\begin{equation}
W = \bQP^2 S + u \bQN^2 S + P \bQP Q + N \bQN Q + S^6.
\label{Csuperpotential;1}
\end{equation}
Under duality, the following $SO(10)$ gauge covariant operators are 
identified
\begin{equation}
\Psi^2_{[{\bf 1}]} \longleftrightarrow P \quad {\rm and} \quad
\Psi^2_{[{\bf 5}]} \longleftrightarrow \bQP^5 \bQN,
\end{equation}
where the subscript $[{\bf n}]$ denotes a rank-$n$ anti-symmetric tensor.

\section{appendix}

The magnetic theory of $Spin(10)$ theory with matter in $F$ vector 
and $N$ spinor representations is summarized. 
As in the appendix A, the irrelevant parameters are set to one. 

The theory is based on an \( [SU(\tN) \times Sp(2N-2)]_{\rm local} 
\times [SU(F) \times SU(2) \times U(1)_Y \times U(1)_R]_{\rm global} \) 
symmetry group. 
Here \( \tN = F+2N-7 \). 
Note that the $SU(N)$ flavor symmetry of the electric theory 
(\ref{AField;N}) is realized as an $SU(2)$ symmetry in this theory. 
The matter content of the theory is as follows,
\begin{eqnarray}
\label{BField;N}
S & \sim & ({\iiBox},{\bf 1};{\bf 1},
{\bf 1},\frac{4F(N-1)}{\tN},2-\frac{2F \xi}{\tN}) 
\nonumber \\
\Qf & \sim & ({\Box},{\bf 1};{\bf 1},
{\bf 2N-1},\frac{2F(N-1)}{\tN} - 2F,2-\frac{F \xi}{\tN}) 
\nonumber \\
\bQF & \sim & ({\bBox},{\bf 1};{\bBox},
{\bf 1},2N - \frac{2F(N-1)}{\tN},\frac{F \xi}{\tN} - \xi)  
\\
\bQA & \sim & ({\bBox},{\bBox};{\bf 1},
{\bf 2},-\frac{2F(N-1)}{\tN},\frac{F \xi}{\tN}) 
\nonumber  \\
R & \sim & ({\bf 1},{\Box};{\bf 1},{\bf 2N-2},2F,0) 
\nonumber  \\
M & \sim & ({\bf 1},{\bf 1};{\iiBox},{\bf 1},4N,2 \xi) 
\nonumber  \\
N & \sim & ({\bf 1},{\bf 1};{\Box},{\bf 2N-1},2N-2F, \xi). 
\nonumber
\end{eqnarray}
The superpotential of the magnetic theory is
\begin{equation}
W = M \bQF^2 S + \bQA^2 S + N \bQF \Qf + \bQA \Qf R.
\label{Bsuperpotential;N}
\end{equation}

\section{appendix}

This appendix provides another consistency checks for the 
superpotential (\ref{W9;22;2}). 
In section 3.1, it is derived as a result of \( SU(7) \to SU(5) \) 
gauge breaking as well as the $Sp(2)$ instanton effect. 

Now let us first take a step backward and consider the 
$SU(7) \times Sp(2)$ gauge theory with superpotential 
$W_{2,2}^{(1)}$ (\ref{W9;22;1}). 
Because the $Sp(2)$ group has 14 flavors, $\bQ_1^{a,\alpha}$ and 
$\bQ_2^{a,\alpha}$ \( (a=1, \cdots ,7) \), it can be dualized by 
$Sp$ duality \cite{fundamental}. 
The new theory has an \( [SU(7) \times Sp(8)]_{\rm local} \times 
SU(2)_{\rm global} \) group and extra dual quarks and mesons 
transforming under the symmetry group as 
\begin{eqnarray}
\label{W22Field;N} 
\bS & \sim & ({\iibBox},{\bf 1};{\bf 1}) 
\nonumber  \\ 
\bA & \sim & ({\tableau{1 1}^{\ast}},{\bf 1};{\bf 3}) 
\\ 
\tQ & \sim & ({\Box},{\Box};{\bf 2}), 
\nonumber 
\end{eqnarray} 
where $SO(9)$ gauge and $SU(F-1)$ flavor symmetries are suppressed. 

The meson $\bS^{ab}$ becomes massive because of its coupling to 
field $S^{ab}$ (see (\ref{W9;22;1})). 
The superpotential of this \( SU(7) \times Sp(8) \) theory is
\begin{eqnarray}
\tW_{2,2}^{(1)} 
& = & \sum_{a,b=1}^{7} \Biggl\{
      S^{8a} \bigl( Q_3^b \bA^{ba}_{3} -  Q_1^b \bA^{ba}_{1} \bigr) +
      S^{88} ( Q_1^a Q_3^b - Q_3^a Q_1^b ) \bA^{ab}_2 +    
      \sum_{i=1,3} \biggl[
      P_i Q_i^a \bQP^a + N_i Q_i^a \bQN^a \biggr] 
\nonumber \\
&   & + \,
      \sum_{\alpha ,\beta =1}^{8} J_{\alpha \beta} \biggl[
      ( \bQP^a \bQP^b + u \bQN^a \bQN^b ) 
      \bigl( \tQ_1^{a,\alpha} \tQ_2^{b,\beta} -
             \tQ_2^{a,\alpha} \tQ_1^{b,\beta} \bigr) +
\nonumber \\
&   & + \,
      \bA^{ab}_1 \tQ_1^{a,\alpha} \tQ_1^{b,\beta} + 
      \bA^{ab}_2 \bigl( \tQ_1^{a,\alpha} \tQ_2^{b,\beta} +
                        \tQ_2^{a,\alpha} \tQ_1^{b,\beta} \bigr) +
      \bA^{ab}_3 \tQ_2^{a,\alpha} \tQ_2^{b,\beta}
      \biggr] \Biggr\} + S^{88}.
\label{tW9;22;1}
\end{eqnarray}

Now, the vanishing F-flatness condition \( \p \tW_{2,2}^{(1)} / 
\p S^{88} \) of this superpotential implies a non-vanishing 
\( \langle Q_1^a Q_3^b \bA^{ab}_2 \rangle \) expectation value. 
This paper takes \( \langle Q_1^7 \rangle \langle Q_3^6 \rangle 
\langle \bA^{67}_2 \rangle \) and concludes that the theory is 
broken to an \( SU(5) \times Sp(8) \) symmetry in which $Sp(8)$ 
gauge group has 10 flavors, \( \tQ_1^{a,\alpha} \) and 
\( \tQ_2^{a,\alpha} \) \( (a=1,\cdots , 5) \). 
Therefore, this $Sp(8)$ theory confines and generates a constraint 
equation on the quantum moduli space, i.e. 
\( {\rm Pf} \tQ^{a,\alpha} = 1 \). 
The result after including this quantum effect is an $SU(5)$ 
gauge theory with the superpotential  
\begin{equation}
\tW_{2,2}^{(2)}  = 
     \sum_{a,b=1}^{5} \Biggl\{
     S^{ab} \bQP^a \bQP^b + u S^{ab} \bQN^a \bQN^b + 
     S^{ab} \bq^a \bq^a + S^{ab} \bA_1^{6a} \bA_3^{7a} +
     \lambda \bigl( 1 - \det S^{ab} \bigr) \Biggr\},
\label{tW9;22;2}
\end{equation}
where \( \bq^a \equiv \bA_1^{7a} = \bA_3^{6a} \), $\lambda$ is the 
Lagrangian multiplier, and \( \det S^{ab} \equiv {\rm Pf} 
\tQ^{a,\alpha} \). 
Note that this superpotential agrees with (\ref{W9;22;2}).

The result of (\ref{W9;21;2}) can also be obtained by first 
dualizing the $Sp(2)$ gauge group of (\ref{W9;21;1}).

\section{appendix}

We briefly report the result when the ``expanded" theory (\ref{W9;N}) 
is perturbed by $p_j$ \( ( 1 \le j \le N-2 ) \). The breaking pattern 
of gauge group is as follows, 
\( SU(2N+4) \times Sp(2N-2) \stackrel{(1)}{\longrightarrow} 
SU(2N+3) \times Sp(2N-2) \stackrel{(2)}{\longrightarrow} 
SU(2N+1) \times Sp(2N-4) \stackrel{(3)}{\longrightarrow} 
\cdots \stackrel{(j)}{\longrightarrow}
SU(2(N-j)+5) \times Sp(2(N-j)) \stackrel{(j+1)}{\longrightarrow} 
SU(2(N-j)+4) \times Sp(2(N-j)-2) \), with superpotential takes 
the same form at breaking stages $(2)$ to $(j)$. 
At the $(j+1)$th-stage, it is an 
\( SU(2(N-j)+4) \times Sp(2(N-j)-2) \) gauge theory with superpotential
\begin{eqnarray}
W_{N,j}^{(j+1)} & = & \sum_{a,b=1}^{2(N-j)+4} \Biggl\{
        S^{ab} \bQP^a \bQP^b + u S^{ab} \bQN^a \bQN^b + 
        S^{ab} \bq^a \bq^b +
        \sum_{i=2j+1}^{2N-1} \biggl[  P_i Q_i^a \bQP^a + 
        N_i Q_i^a \bQN^a + h_i Q_i^a \bq^a \biggr]  \nonumber \\
  &  & + \sum_{\alpha ,\beta =1}^{2(N-j)-2} J_{\alpha \beta} \biggl[
       S^{ab} \big( \bQ_1^{a,\alpha} \bQ_2^{b,\beta} -
              \bQ_2^{a,\alpha} \bQ_1^{b,\beta} \big) +
       \sum_{i=2j+1}^{2N-1} Q_i^a 
              \big( \bQ_1^{a,\alpha} R_i^\beta + 
              \bQ_2^{a,\alpha} R_{i-1}^\beta \big) 
       \biggr] \nonumber \\
  &  & + \sum_{i=2j+1}^{2N-1} \biggl[  R_i^{2(N-j)-1} Q_i^a \bq^a + 
        R_i^{2(N-j)} Q_i^a \bQ_1^{a,2(N-j)-1} \biggr] \Biggr\} \, ,
\label{W9;Nj;j+1}
\end{eqnarray}
where \( \bq^a \equiv \bQ_1^{a,2N-2} = \bQ_2^{a,2N-3} \). 
By F-flatness conditions, \( R_i^{2(N-j)-1} = J_{\alpha \beta} 
Q_{2j+1}^b \bQ_2^{b,\alpha} R_{i+1}^\beta \) and 
\( \bQ_1^{a,2(N-j)-1} = J_{\alpha \beta} Q_{2j+1}^b \bQ_2^{b,\alpha} 
\bQ_2^{a,\beta} \) are expressed as composite states of the 
\( SU(2(N-j)+4) \times Sp(2(N-j)-2) \) gauge group. 
However, the $Sp(2(N-j)-2)$ gauge group is in IR free phase, 
these expressions can be fulfilled only if the $Sp(2(N-j)-2)$ group 
is broken to a smaller one.

Because no further Higgs mechanism occurs in (\ref{W9;Nj;j+1}), 
\( R_i^{2(N-j)-1} \) and \( \bQ_1^{a,2(N-j)-1} \) vanish. 
We thus conclude \( W_{N,j}^{(j+1)} = W_{N-j,0} \), where ``0" 
subscript denotes the elimination of $p_j$ in (\ref{W9;N}).



\newpage

\begin{thebibliography}{100}

\bibitem{susy} 
N. Seiberg, {\it Phys.\ Rev.\ }{\bf D49} (1994) 6857, 
{\tt hep-th/9402044};
K. Intriligator, R. G. Leigh and N. Seiberg, 
{\it Phys.\ Rev.\ }{\bf D50} (1994) 1092, {\tt hep-th/9403198};
K. Intriligator, 
{\it Phys.\ Lett.\ }{\bf B336} (1994) 409, {\tt hep-th/9407106};
O. Aharony, 
{\it Phys.\ Lett.\ }{\bf B351} (1995) 220, {\tt hep-th/9502013}.

\bibitem{recent} 
N. Seiberg, 
{\tt hep-th/9408013},{\tt hep-th/9506077}; 
K. Intriligator and N. Seiberg, 
{\it Nucl.\ Phys.\ Proc.\ Suppl.\ }{\bf 45BC} (1996) 1, 
{\tt hep-th/9509066}.

\bibitem{seiberg}
N. Seiberg,
{\it Nucl.\ Phys.\ }{\bf B435} (1995) 129, {\tt hep-th/9411149}.

\bibitem{fundamental} 
K. Intriligator and N. Seiberg, 
{\it Nucl.\ Phys.\ }{\bf B444} (1995) 125, {\tt hep-th/9503179};
K. Intriligator and P. Pouliot, 
{\it Phys.\ Lett.\ }{\bf B353} (1995) 471, {\tt hep-th/9505005}. 

\bibitem{many1} 
D. Kutasov,
{\it Phys.\ Lett.\ }{\bf B351} (1995) 230, {\tt hep-th/9503086}; 
D. Kutasov and A. Schwimmer,
{\it Phys.\ Lett.\ }{\bf B354} (1995) 315, {\tt hep-th/9505004}; 
D. Kutasov, A. Schwimmer and N. Seiberg, 
{\it Nucl.\ Phys.\ }{\bf B459} (1996) 445, {\tt hep-th/9510222}. 

\bibitem{many2}
R. G. Leigh and M. J. Strassler, 
{\it Nucl.\ Phys.\ }{\bf B447} (1995) 95, {\tt hep-th/9503121};
O. Aharony, J. Sonnenschein and S. Yankielowicz, 
{\it Nucl.\ Phys.\ }{\bf B449} (1995) 509, {\tt hep-th/9504113};
K. Intriligator, 
{\it Nucl.\ Phys.\ }{\bf B448} (1995) 187, {\tt hep-th/9505051}; 
R. G. Leigh and M. J. Strassler, 
{\it Phys.\ Lett.\ }{\bf B356} (1995) 492, {\tt hep-th/9505088};
I. Pesando, 
{\it Mod.\ Phys.\ Lett.\ }{\bf A10} (1995) 1871, {\tt hep-th/9506139};
K. Intriligator, R. G. Leigh and M. J. Strassler, 
{\it Nucl.\ Phys.\ }{\bf B456} (1995) 567, {\tt hep-th/9506148}; 
S. B. Giddings and J. M. Pierre, 
{\it Phys.\ Rev.\ }{\bf D52} (1995) 6065, {\tt hep-th/9509196};  
E. Poppitz, Y. Shadmi and S.P. Trievi, 
{\it Nucl.\ Phys.\ }{\bf B480} (1996) 125, {\tt hep-th/9605113}, 
{\it Phys.\ Lett.\ }{\bf B388} (1996) 561, {\tt hep-th/9606184};
J. H. Brodie,
{\it Nucl.\ Phys.\ }{\bf B478} (1996) 123, {\tt hep-th/9605232};
C. Csaki, W. Skiba, and M. Schmaltz,
{\it Nucl.\ Phys.\ }{\bf B487} (1997) 128, {\tt hep-th/9607210};
P. Ramond, 
{\tt hep-th/9608077};
N. Maru and S. Kitakado, 
{\tt hep-th/9609230}; 
J. H. Brodie and M. J. Strassler, 
{\tt hep-th/9611197}.

\bibitem{chi-nonchi} 
P. Pouliot, 
{\it Phys.\ Lett.\ }{\bf B359} (1995) 108, {\tt hep-th/9507018}; 
P. Pouliot and M. J. Strassler, 
{\it Phys.\ Lett.\ }{\bf B370} (1996) 76, {\tt hep-th/9510228}; 
P. Pouliot and M. J. Strassler, 
{\it Phys.\ Lett.\ }{\bf B375} (1996) 175, {\tt hep-th/9602031}; 
T. Kawano, 
{\it Prog.\ Theor.\ Phys.\ }{\bf 95} (1996) 963, {\tt hep-th/9602035};
C. Csaki, W. Skiba, and M. Schmaltz,
{\it Nucl.\ Phys.\ }{\bf B487} (1997) 128, {\tt hep-th/9607210};
P. Cho,
{\tt hep-th/9702059}.

\bibitem{spin10}
M. Berkooz, P. Cho, P. Kraus, and M. Strassler, 
{\tt hep-th/9705003}.

\bibitem{spinor-monopole}
M. Strassler,
{\tt hep-th/9709081}.

\bibitem{weinberg}
E. J. Weinberg, D. London, and J. L. Rosner,
{\it Nucl.\ Phys.\ }{\bf 236} (1984) 90.

\bibitem{accidental}
R. G. Leigh and M. J. Strassler, 
{\it Nucl.\ Phys.\ }{\bf B496} (1997) 132, {\tt hep-th/9611020};
J. Distler and A. Karch,
{\tt hep-th/9611088}.

\bibitem{decon1}  
M. Berkooz, 
{\it Nucl.\ Phys.\ }{\bf B452} (1995) 513, {\tt hep-th/9505067}; 
P. Pouliot, 
{\it Phys.\ Lett.\ }{\bf B367} (1996) 151, {\tt hep-th/9510148};
M. A. Luty, M. Schmaltz and J. Terning, 
{\it Phys.\ Rev.\ }{\bf D54} (1996) 7815, {\tt hep-th/9603034};
T. Sakai, 
{\tt hep-th/9701155}.

\bibitem{decon2}
W.-C. Su, {\tt hep-th/9707076}.

\end{thebibliography}
\end{document}